\documentstyle[12pt]{article}
\begin{document}
\setlength{\oddsidemargin}{-.5cm}
\setlength{\topmargin}{-2.cm}

\Large \begin{center}
Combining multigrid and wavelet ideas to construct 
more efficient iterative multiscale algorithms
\end{center} 

\large \begin{center}
Stefan Goedecker, Claire Chauvin
\end{center} 
\normalsize \begin{center}
D\'epartement de recherche fondamentale sur la mati\`ere condens\'ee,\\
         SP2M/L\_Sim, CEA-Grenoble, 38054 Grenoble cedex~9, France
\end{center} 

\section{Abstract}
It is shown how various ideas that are well established for the solution 
of Poisson's equation using plane wave and multigrid methods 
can be combined with wavelet concepts. 
The combination of wavelet concepts and multigrid techniques 
turns out to be particularly fruitful.
We propose a modified multigrid V cycle scheme that is not only much 
simpler, but also more efficient than the standard 
V cycle. Whereas in the traditional V cycle the residue is 
passed to the coarser grid levels, this new scheme does not require 
the calculation of a residue. Instead it works with copies of 
the charge density on the different grid levels that were obtained 
from the underlying charge density on the finest grid by wavelet 
transformations. This scheme is not limited to the pure wavelet 
setting, where it is faster than the preconditioned conjugate gradient method, 
but equally well applicable for finite difference discretizations.  

\section{Introduction}
Poisson's equation and Schr\"{o}dinger's equation are the central equations 
for atomistic simulations. In case force fields \cite{field} are used, 
Poisson's equation handles the long range electrostatic interactions. 
If the forces are calculated quantum mechanically, electronic structure 
calculations have to be performed. Selfconsistent electronic structure 
calculations require the solution of a system where Schr\"{o}dinger's 
equation is coupled with Poisson's equation. Multiscale approaches for 
the solution of these two central equations are widely used, since they 
are much more efficient for big system sizes than traditional 
approaches. 

The multigrid method has been used for the solution of Poisson's 
equation in the context of classical molecular dynamics simulations
\cite{sagui} and it has been used by various groups for electronic 
structure calculations. Several flavors of multigrid for 
electronic structure calculations have been proposed: Its direct 
solution by the Full Approximation scheme \cite{beck}, the 
Rayleigh Quotient Multigrid method \cite{nieminen} and a scheme 
where the the solution of the linear system of equations arising 
from the preconditioning step is performed by multigrid 
\cite{briggs,fattebert}. The linear system solved in the preconditioning 
steps is not Schr\"{o}dinger's equation, but Poisson's equation. 
The reason for this is that it is too difficult to find 
coarse grained representations of the Hamiltonian operator if 
nonlocal pseudopotentials are used and the deferred defect correction 
scheme \cite{deferred} justifies the replacement of the Hamiltonian by 
the Laplacian. Thus, it turns out that in all cases it is 
Poisson's equation that has to be solved using multigrid. 

Another very promising multiscale approach to electronic structure 
calculations is the use of wavelets as basis sets \cite{ariasrmp}.
As with any large systematic basis set it is also in the 
wavelet context very important to use efficient preconditioning 
schemes. Diagonal preconditioning \cite{prec} is most widely used.
However it would be desirable to have at our disposal more 
powerful preconditioning schemes. Using multigrid ideas for 
preconditioning has already been proposed in the context of 
interpolating wavelets \cite{preprint}. Our discussion of 
prolongation and restriction schemes based on wavelet theory 
will show that the class of schemes based on interpolating wavelets 
is not optimal. 

Because of its central importance for atomistic simulations and 
because it is a prototype equation for the study of new methods 
we will from now on consider only 
Poisson's equation 
\begin{equation} \label{poisson} 
\nabla^2 V({\bf r}) = - 4 \pi \rho({\bf r}) 
\end{equation}
The solution of the differential equation Eq.~\ref{poisson} 
can be written as an integral equation
\begin{equation} \label{inteq} 
 V({\bf r}) = \int \frac{\rho({\bf r}') }{|{\bf r} - {\bf r}'|}
\end{equation}
Gravitational problems are based on exactly the same equations as 
the electrostatic problem, but we will use in this article the language of 
electrostatics, i.e. we will refer to $\rho({\bf r})$ as a charge density.
The most efficient numerical approaches for the solution of electrostatic 
problems are based on Eq~\ref{poisson} rather than Eq.~\ref{inteq}. However 
preconditioning steps found in these methods can be considered as approximate 
solutions of Eq.~\ref{inteq}.
The fact that the Green's function $\frac{ 1 }{|{\bf r} - {\bf r}'|}$ is 
of long range makes the numerical solution of Poisson's equation 
difficult, since it 
implies that a charge density at a point ${\bf r}'$ will have an non-negligible 
influence on the potential $V({\bf r})$ at a point ${\bf r}$ far away.
A naive implementation of Eq.~\ref{inteq} would therefore have 
a quadratic scaling. 
It comes however to our help, that the potential arising from a charge 
distribution far away is slowly varying and does not depend on the 
details of the charge distribution. 
All efficient algorithms for solving electrostatic problems 
are therefore based on a hierarchical multiscale treatment. On the short 
length scales the rapid variations of the potential due to the exact 
charge distribution of close by sources of 
charge are treated, on the large length scales the slow variation due to 
some smoothed charge distribution of far sources is accounted for. 
Since the number of degrees of freedom decreases rapidly with increasing 
length scales, one can obtain algorithms with linear or nearly linear scaling. 
In the following, we will briefly summarize how this hierarchical treatment 
is implemented in the standard algorithms

\begin{itemize}
\item Fourier Analysis: \newline
If the charge density is written in its Fourier representation 
$$ \rho({\bf r})  = \sum_{\bf K} \rho_{\bf K} e^{i {\bf K} {\bf r}} $$
the different length scales that are in this case given by 
$ \lambda = \frac{2 \pi}{K}$ decouple entirely and the Fourier representation 
of the potential is given by 
\begin{equation} \label{fourpot}
 V({\bf r})  = \sum_{\bf K} \frac{\rho_{\bf K}}{K^2} e^{i {\bf K} {\bf r}} 
\end{equation}
The Fourier analysis of the real space charge density necessary to obtain 
its Fourier components $\rho_{\bf K}$ and the synthesis of the 
potential in real space from its Fourier components given by Eq.~\ref{fourpot} can 
be done with Fast Fourier methods at a cost of $N \: log_2(N)$ where N is the 
number of grid points. The solution of Poisson's equation in a plane 
wave is thus a divide and conquer approach where the division is into 
the single Fourier components.

\item Multigrid methods (MG): \newline
Trying to solve Poisson's equation by any relaxation or iterative 
method (such as conjugate gradient) on the fine grid on which one 
finally wants to have the solution leads to a number of iterations that 
increases strongly with the size of the grid. The reason for this is that on a 
grid with a given spacing $h$ one can efficiently treat Fourier components 
with a wavelength $\lambda = \frac{2 \pi}{K}$ that is comparable to 
the the grid spacing $h$, but the longer wavelength Fourier components 
converge very slowly. This increase in the number of iterations prevents 
a straightforward linear scaling solution of Eq.~\ref{poisson}. 
In the multigrid method, pioneered by A. Brandt~\cite{brandt},  
one is therefore introducing a hierarchy of grids with a grid spacing that 
is increasing by a factor of two on each hierarchic level. In contrast to the 
Fourier method where the charge and the potential are directly decomposed 
into components characterized by a certain length scale, it is the residue 
that is passed from the fine grids to the coarse grids in the MG method. 
The residue corresponds to the charge that would give rise to a potential 
that is the difference between the exact potential and the approximate 
potential at the current stage of the iteration. 

\end{itemize}

The solution of partial differential equations in a wavelet basis is typically 
done by preconditioned iterative techniques~\cite{pde}. The diagonal preconditioning 
approach, that is based on well established plane wave techniques, will be presented 
in the next section. The section after the next will introduce multigrid for 
Poisson's equation in a wavelet basis. Even though the fundamental similarities 
between wavelet and multigrid schemes have been recognized by many workers 
(such as in ref.~\cite{review}) this sections contains to the best of our knowledge the 
first thorough discussion of how both methods can profit from each other.

Within wavelet theory~\cite{daub} one has two possible representations 
of a function $f(x)$, a scaling function representation
\begin{equation} 
 f(x) = \sum_j s_j^{Lmax} \phi_j^{Lmax}(x) \label{scfrep} 
\end{equation}
and a wavelet representation.
\begin{equation} 
f(x) = \sum_j s_j^{Lmin} \: \phi_j^{Lmin}(x) +
       \sum_{l=Lmin}^{Lmax} \sum_j d_j^{l} \; \psi_j^{l}(x)  \label{wvltrep} \: .
\end{equation}
In contrast to the scaling function representation, the wavelet 
representation is a hierarchic representation. The wavelet at the 
hierarchic level $l$ is related to the mother wavelet $\psi$ by 
\begin{equation} \label{scalrel}
\psi_{i}^{l}(x)   =  \sqrt{2}^l \psi(2^l x-i)
\end{equation}
The characteristic length scale of a wavelet at resolution 
level $l$ is therefore proportional to $2^{-l}$.  
A wavelet on a certain level $l$ is a linear combination of 
scaling functions at the higher resolution level $l+1$
\begin{equation} 
\psi_{i}^{l}(x) = \sum_{j=-m}^{m} g_j \: \phi_{2 i+j}^{l+1}(x)
\end{equation}
Scaling functions at adjacent resolution levels are related by 
a similar refinement relation 
\begin{equation} 
\phi_{i}^{l}(x) = \sum_{j=-m}^{m} h_j \: \phi_{2 i+j}^{l+1}(x)
\end{equation}
and hence also any wavelet at a resolution level $l$ 
is a linear combination of the highest resolution scaling functions.
The so-called fast wavelet transform allows us to transform back 
and forth between a scaling function and a wavelet representation. 

Let us now introduce wavelet representations of the potential and the 
charge density
\begin{eqnarray} 
V(x) & = & \sum_j V_j^{Lmin} \: \phi_j^{Lmin}(x) +
       \sum_{l=Lmin}^{Lmax} \sum_j V_j^{l} \; \psi_j^{l}(x)  \label{potrep} \\ 
\rho(x) & = & \sum_j \rho_j^{Lmin} \: \phi_j^{Lmin}(x) +
       \sum_{l=Lmin}^{Lmax} \sum_j \rho_j^{l} \; \psi_j^{l}(x)  \label{rhorep}
\end{eqnarray}

Different levels do not completely decouple, i.e the components on 
level $l$, $V_j^l$, of the exact overall solution do not satisfy the 
single level Poisson equation 
\begin{equation} 
\nabla^2 \left( \sum_j V_j^{l} \; \psi_j^{l}(x) \right) \neq 
 - 4 \pi \left( \sum_j \rho_j^{l} \; \psi_j^{l}(x) \right)
\end{equation}
within the chosen discretization scheme. 
This is due to the fact that the wavelets are not perfectly localized 
in Fourier space, i.e. many frequencies are necessary to synthesize 
a wavelet. However the amplitude of all these frequencies is clearly peaked 
at a nonzero characteristic frequency for any wavelet with at least 
one vanishing moment. From the scaling property (Eq.~\ref{scalrel}) it 
follows, that the frequency at which the peak occurs changes by a factor 
of two on neighboring resolution grids. This suggest that 
the coupling between the different resolution levels is weak.  

In the preceding paragraph we presented the mathematical framework only 
for the one-dimensional case. The generalization to the 3-dim case 
is straightforward by using tensor products~\cite{daub}. 
Also in the rest of the paper only the one-dimensional form of the 
mathematical formulas will presented for reasons of simplicity. 
It has to be stressed however that all the numerical results were 
obtained for the three-dimensional case and with periodic boundary 
conditions.

\section{The diagonal preconditioning approach for wavelets}
Preconditioning requires finding a matrix with a simple structure 
that has eigenvalues and eigenvectors that are similar to the ones 
of the matrix in question~\cite{dahmen,prec}. The structure has to be simple in the sense 
that it allows us to calculate the inverse easily. The simplest and 
most widely used structure in this respect is the structure of a 
diagonal matrix. As will be shown a diagonal preconditioning matrix 
can be found in a wavelet basis set and preconditioned conjugate 
gradient type methods are then a possible method
for the solution of Poisson's equation expressed in
differential form (Eq.\ref{poisson}).

As one adds successive levels of wavelets to the basis set the largest 
eigenvalue grows by a factor of 4 for each level. This can easily 
be understood from Fourier analysis. As one increases the resolution 
by a factor of 2 (i.e. increases the largest Fourier vector $k_{max}$ 
by a factor of 2) the largest eigenvalue increases by a factor of 4.
This basic scaling property of the eigenvalue spectrum can easily 
be modeled by a diagonal matrix, where all the diagonal elements 
are all equal on one resolution level, but increase by a factor of 
4 as one goes to a higher resolution level. It is of course clear that 
all the details of the true spectrum are not reproduced by this 
approximations. The true spectrum consists of a large number 
of moderately degenerate eigenvalues. The spectrum of the approximate 
matrix consists of a few highly degenerate eigenvalues where 
each eigenvalue has all the scaling functions of one resolution 
level as its eigenfunctions. The true eigenfunctions are of course 
mixtures of scaling functions on different resolution levels, but if 
the wavelet family is well localized in Fourier space the contributions 
from neighboring resolution levels are weak. The localization in Fourier 
space increases with the number of vanishing moments~\cite{mywvltbook} and 
therefore this diagonal preconditioning method works for instance much 
better for lifted interpolating wavelets~\cite{schneider,sweldens} 
than for ordinary interpolating~\cite{lazy} 
wavelets~\cite{mycip}.

\begin{figure}[p]
     \begin{center}
      \setlength{\unitlength}{1cm}
       \begin{picture}( 5.,18.)           
        \put(-5., 8.){\includegraphics{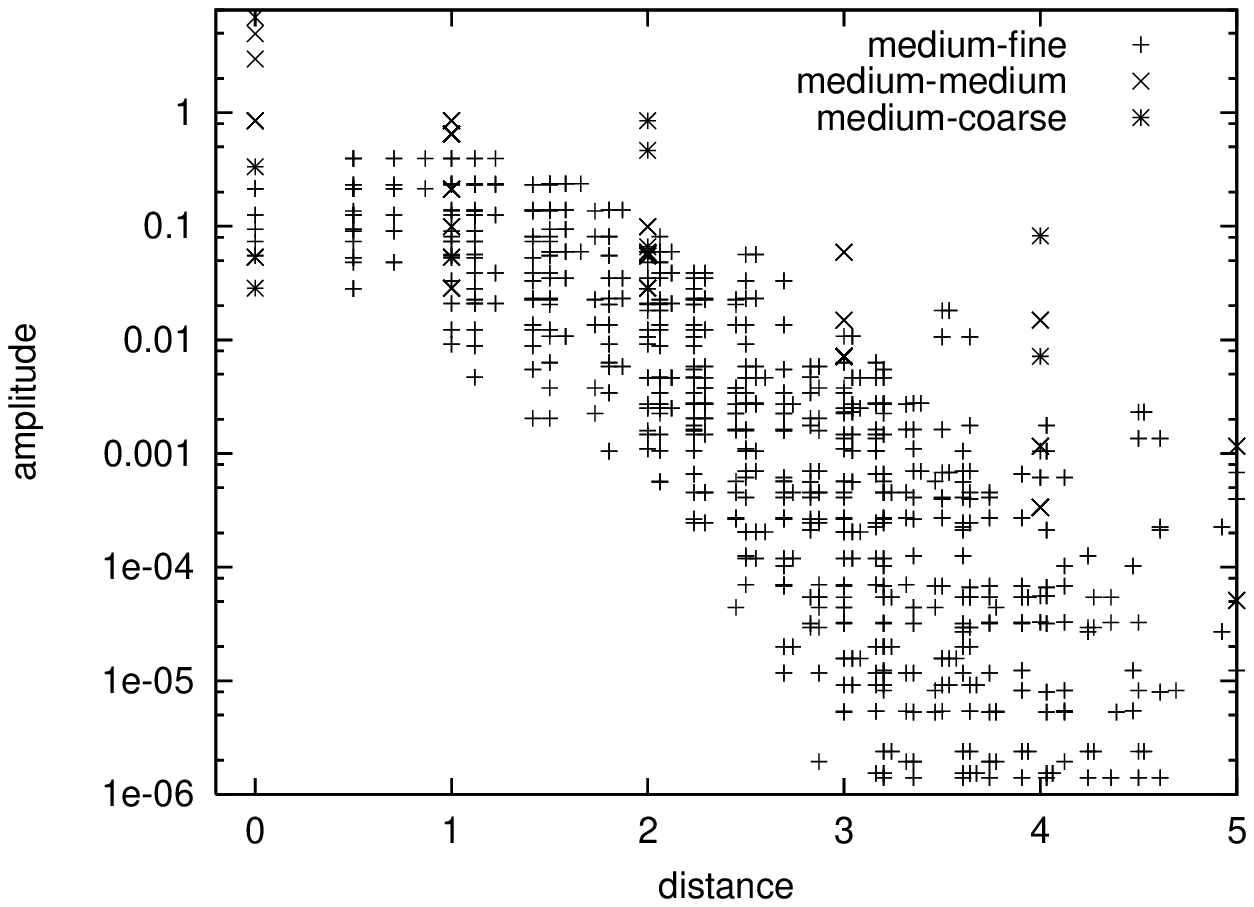}}   
        \put(-5.,-1.5){\includegraphics{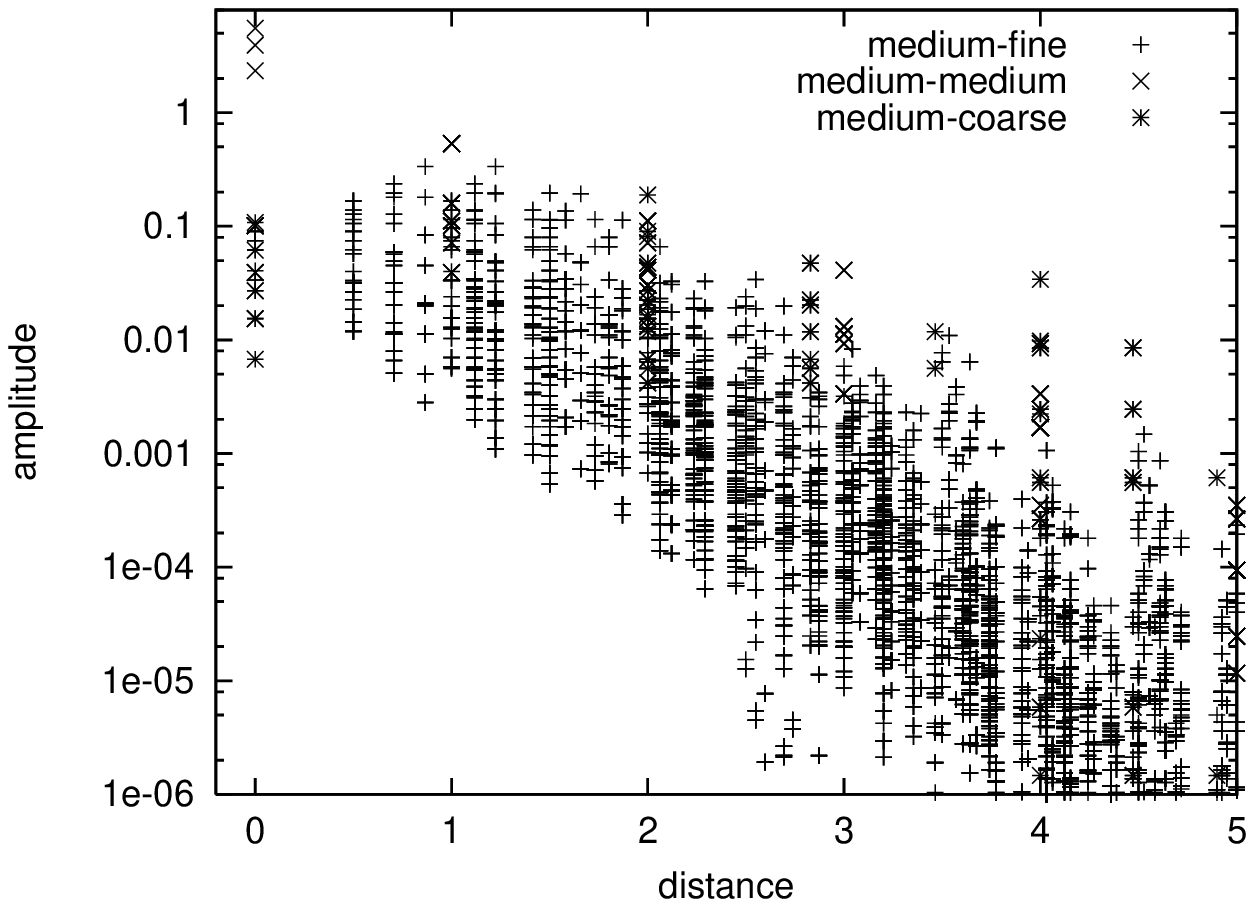}}   
       \end{picture}
 \caption[]{\label{decay} The absolute value of the amplitude of the matrix 
elements of the Laplacian in a basis of 6-th order interpolating 
(top panel) and 6-th order lifted interpolating (lower panel) wavelets. 
Shown are the elements within one resolution block (medium-medium) as well 
as the elements coupling to a higher resolution (medium-fine) and a lower 
resolution level (medium-coarse). The distance 1 corresponds to the nearest 
neighbor distance on the medium resolution level. Because of the 
better localization in Fourier space of the lifted wavelets, the matrix 
is more diagonally dominant in the lifted wavelet representation.  }
      \end{center}
     \end{figure}

Another line of arguments, that shows the weakness of the diagonal 
preconditioning, is the following. The preconditioning matrix can also be considered to be 
the diagonal part of the matrix representing the Laplacian~\cite{beylkin}. Since the 
diagonal elements increase by a factor of 4 on each higher resolution level,
\begin{eqnarray} \label{factor4}
\int \tilde{\psi}_0^{l+1}(x) \frac{\partial^2}{\partial x^2} \psi_0^{l+1}(x) dx & = & 
2 \int \tilde{\psi}_0^{l}(2 x ) \frac{\partial^2}{\partial x^2} 
        \psi_0^{l}(2 x ) dx \\  & = & 
4 \int \tilde{\psi}_j^{l}(x) \frac{\partial^2}{\partial x^2} \psi_j^{l}(x) dx  
\nonumber
\end{eqnarray}
the spectrum of the matrix has again the correct scaling properties. 
Eq.~\ref{factor4} involves both the wavelets $\psi$ and their duals $\tilde{\psi}$
since it was written down for the most general context of a Petrov-Galerkin 
approach in a biorthogonal basis. In a pure Galerkin context or for 
orthogonal wavelet families the $\tilde{\psi}$ have to be replaced by the $\psi$'s.
In the 3-dimensional case there are three different types of wavelets 
(products of 2 scaling functions and 1 wavelet, 
 products of 1 scaling function and 2 wavelets and 
products of 3 wavelets). Each type of wavelet gives rise to a different 
diagonal element, but again all these diagonal elements differ by 
a factor of 4 on different resolution levels.
Because of the weak coupling between different resolution levels 
discussed above, we expect the matrix elements of the Laplacian 
involving wavelets at two different resolutions levels to be small.
The numerical examination of the matrix elements 
(Fig.~\ref{decay}) confirms this guess. It also shows that 
within one resolution level the amplitude of the matrix elements 
decays rapidly with respect to the distance of the two wavelets 
and is zero as soon as they do not any more overlap. 
Nevertheless, some matrix 
elements coupling nearest neighbor wavelets are not much smaller 
than the diagonal elements. One also finds a few matrix elements 
between different resolution levels that are less than 
one order of magnitude smaller than the one within a single resolution level. 

The fact that all off-diagonal matrix elements are neglected in 
current precondition schemes explains their relatively slow convergence. 
It amounts to finding an approximate Greens function that is 
diagonal in a wavelet representation. This is obviously a 
rather poor approximation. Let us nevertheless stress that this diagonal 
matrix obtained by inverting a diagonal approximation to the Laplacian is 
a much more reasonable approximation for preconditioning purposes 
than the diagonal part of the Greens function. 
The diagonal part of the Greens function has actually completely 
different scaling properties. The elements increase by a factor of 2 as 
one goes to higher resolution levels instead of decreasing by a factor of 4.
The multigrid methods to be discussed 
later include also in an approximative way through Gauss-Seidel 
relaxations this off-diagonal coupling within each resolution block as 
well as the coupling between the different resolutions levels.

In the following we will present some numerical results for the solution 
of the 3-dimensional Poisson equation in a wavelet basis using the 
diagonal preconditioning approach. 
All the methods presented in this paper will have the property that 
the convergence rate is independent of the grid size.
We have chosen $64^3$ grids for all the numerical examples.
The fact that the number of iterations necessary to reach a certain 
target accuracy is independent of the system size together with the 
fact that a single iteration involves a cost that is linear with 
respect to the number of grid points ensures that the Poisson's 
equation can be solved with overall linear scaling. Whereas we use here 
only simple equidistant grids, this linear scaling has already been 
demonstrated with highly adaptive grids in problems that involve 
many different length scales~\cite{mycip,myscc,ariasjcp,ariasrmp}.

The preconditioning step using simply the diagonal is given by
\begin{equation} \label{pwprec}
 \Delta V_j^l = const \: \: 4^{-l} \Delta \rho_j^l
\end{equation}
In analogy to Eq.~\ref{potrep},\ref{rhorep}, 
the $\Delta \rho_j^l$'s are the wavelet coefficients on the 
$l$-th resolution level of the residue 
$\Delta \rho({\bf r}) = \nabla^2 \tilde{V}({\bf r})  + 4 \pi \rho({\bf r}) $ 
in a wavelet representation. $\tilde{V}({\bf r})$ 
is the approximate solution at a certain iteration of the solution 
process. The preconditioned 
residue $\Delta V$ is then used to update the approximate potential $\tilde{V}$.
In the case of the preconditioned steepest descent method used here 
this update simply reads  
\begin{equation} \label{stdes}
 \tilde{V} \leftarrow \tilde{V} + \alpha \Delta V
\end{equation}
where $\alpha$ is an appropriate step size. 
As discussed above the constant in Eq.~\ref{pwprec} depends in the 
three dimensional case upon which type of wavelet is implied since it is 
the inverse of the Laplace matrix element between two wavelets of this type.

\begin{figure}[ht]
     \begin{center}
      \setlength{\unitlength}{1cm}
       \begin{picture}( 5.,6.)           
        \put(-7.,-1.5){\includegraphics{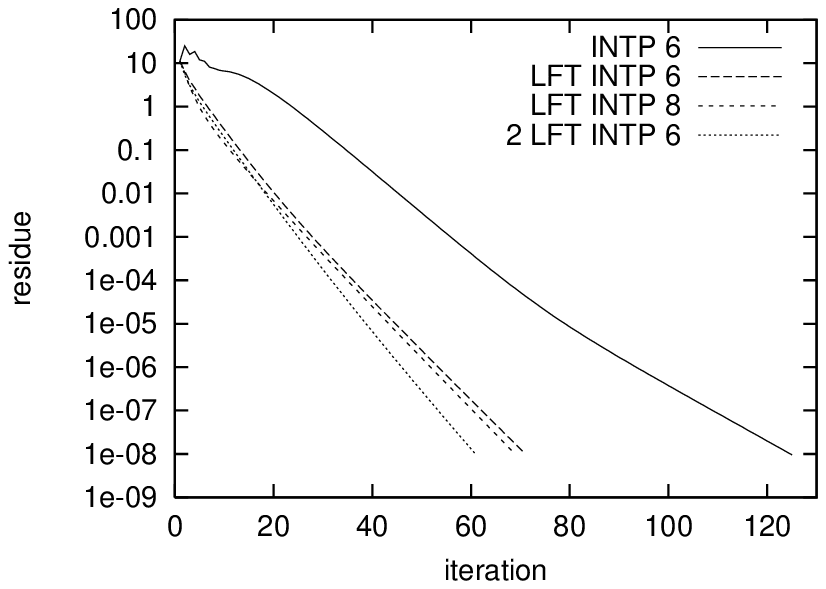}}   
        \put(1.,-1.5){\includegraphics{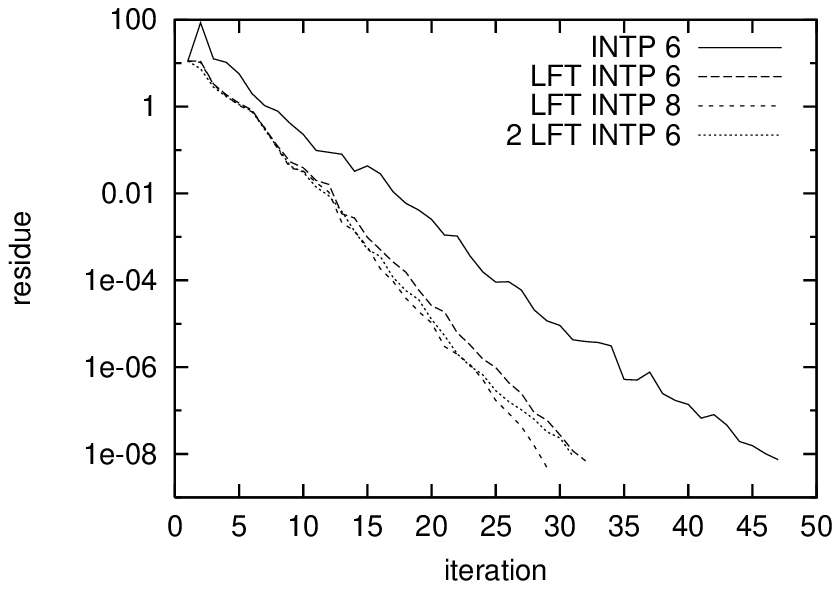}}   
       \end{picture}
 \caption[]{\label{sd_pw}
The reduction of the residue 
during a steepest descent iteration (left hand panel) 
and a FGMRES iteration (right hand panel) with interpolating and lifted 
interpolating wavelets. } 
      \end{center}
     \end{figure}

Fig.~\ref{sd_pw} shows numerical results for several wavelet families.
The slow convergence of the interpolating wavelets is due to the fact that 
they have a non-vanishing average and therefore a non-vanishing 
zero Fourier component~\cite{mycip}. 
Hence they are all localized in Fourier space 
at the origin instead of being localized around a non-zero frequency.
This deficiency can be eliminated by lifting. The Fourier power 
spectrum of the lifted wavelets tends to zero at the origin with 
zero slope for the family with two vanishing moments considered here.
Lifting the wavelet twice leads to 4 vanishing moments and a even better 
localization in Fourier space. The improvement in the convergence rate 
is however only marginal. 
The higher 8-th order lifted interpolating wavelet is smoother than 
its 6-th order counterpart and hence better localized in the 
high frequency part. This also leads to a slightly faster convergence.

Combining the diagonal preconditioning (Eq.~\ref{pwprec}) with a 
FGMRES convergence accelerator~\cite{diis} instead of using it 
within a steepest descent minimization 
gives a significantly faster convergence. The number of iterations can 
nearly be cut into half as shown in Fig~\ref{sd_pw}. 

\begin{figure}[ht]
     \begin{center}
      \setlength{\unitlength}{1cm}
       \begin{picture}( 5.,6.)           
        \put(-5.,-1.5){\includegraphics{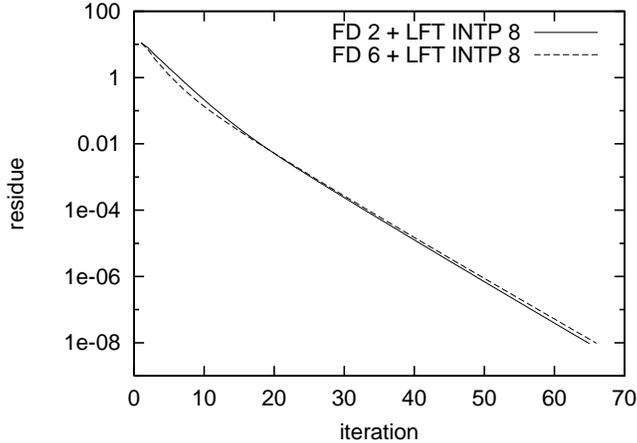}}   
       \end{picture}
 \caption[]{\label{pw_fd} The convergence rate for the case where Poisson's 
 equation is solved with finite differences and 8-th order lifted wavelets 
are used for the preconditioned steepest descent.}
      \end{center}
     \end{figure}

Up to now we have only considered the case where the elements of the 
matrix representing the Laplacian were calculated within the same wavelet 
family that was used to analyze the residue by wavelet transformations 
to do the preconditioning step.
More general schemes can however be implemented. It is not even necessary 
that the calculation of the Laplacian matrix elements is done in a 
wavelet basis. One can instead use simple second order finite differences, 
which in the one-dimensional case are given by 
\begin{equation} \label{fdlow}
\frac{1}{h^2} (-V_{i-1}  + 2 V_i - V_{i+1}) \:, 
\end{equation}
or some higher order finite differences for the 
calculation of the matrix elements. The scaling relation Eq.~\ref{factor4} 
does not any more hold exactly, but it is fulfilled approximately and 
the schemes works as well as in the pure wavelet case 
as is shown in Fig.~\ref{pw_fd}.

\section{The MG approach for wavelets}
The aim of this part of the article is twofold. One aspect is how to 
speed up the convergence of the solution process for Poisson's equation 
expressed in a wavelet basis set compared to the diagonal preconditioning approach. 
The other aspect is 
how to accelerate multigrid schemes by incorporating wavelet concepts.
The part therefore begins with a brief review of the multigrid method.

\begin{figure}[ht]
     \begin{center}
      \setlength{\unitlength}{1cm}
       \begin{picture}( 5.,9.)           
        \put(-8.,-9.5){\includegraphics{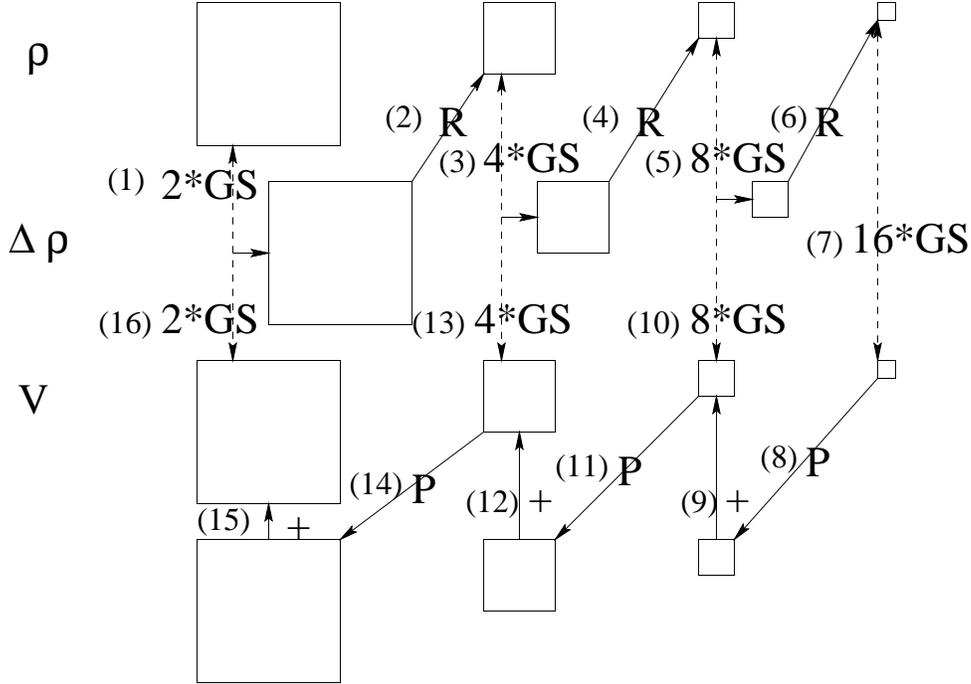}}   
       \end{picture}
 \caption[]{\label{mgv} Schematic representation of a multigrid V 
cycle as described in the text. GS denotes a red-black Gauss-Seidel 
relaxation, R restriction, P prolongation and + addition of the data 
sets. The numbering in parentheses gives the ordering of the different steps 
of the algorithm. }
      \end{center}
     \end{figure}

Fig.~\ref{mgv} schematically shows the algorithm of a standard  multigrid 
V cycle~\cite{trottenberg,textbooks}. 
Even though the scheme is valid in any dimension, a two dimensional 
setting is suggested by the figure, since the data is represented as 
squares. Since less data is available on the coarse grids, the squares 
holding the coarse grid data are increasingly smaller. It has to be stressed 
that the remarks of the end of the first part remain valid and that in 
particular all the numerical calculations are 3-dim calculations. The upper 
half of the  figure shows the first part of the V cycle where one goes from the 
finest grid to the coarsest grid and the lower half the second part where one 
goes back to the finest grid. 

In the first part of the V cycle the 
potential on all hierarchic grids is improved by a standard 
red-black Gauss-Seidel relaxation denoted by GS. The GS relaxation reduces 
the error components of wavelengths $\lambda$ that are comparable to the grid 
spacing $h$ very efficiently. In the  3-dimensional case we are considering here, 
the smoothing factor is .445 (page 74 of ref~\cite{trottenberg}). 
Since we use 2 GS 
relaxations roughly one quarter of the error around the wavelength $h$ 
survives the relaxations on each level. 
As a consequence the residue $\Delta \rho$ 
contains mainly longer wavelengths which then in turn are again efficiently 
eliminated by the GS relaxations on the coarser grids. Nevertheless, the 
remaining quarter of the shorter wavelengths surviving the 
relaxations on the finer grid pollutes the 
residue on the coarser grid through aliasing effects. 
Aliasing pollution means that 
even if the residue on the finer grid would contain only 
wavelengths around $h$ (and in particular no wavelength around $2h$) 
the restricted quantity would not be identically zero.  

In the second part of the V cycle the solutions obtained by relaxation 
on the coarse grid are prolongated to the fine grids and added to the 
existing solutions on each level. Aliasing pollution is again present in the 
prolongation procedure. Due to the accumulated aliasing errors 2 GS 
relaxations are again done on each level before proceeding to the next 
finer level.

To a first approximation the different representations of $\rho$ at the top 
of Fig.~\ref{mgv} represent Fourier filtered versions of the real 
space data set $\rho$ on the finest grid. The large 
data set contains all the Fourier components, while the smaller data 
sets contain only lower and lower frequency parts of $\rho$. In the 
1-dimensional case only the lower half of the spectrum is still 
dominating when going to the coarse grid, in the 3-dimensional case 
it is only one eight of the spectrum. Because 
of the various aliasing errors described above the Fourier decomposition 
is however not perfect. Obviously it would be desirable to make 
this Fourier decomposition as perfect as possible. In the absence of 
aliasing errors, the GS 
relaxations would not have to deal with any Fourier components spilling 
over from higher or lower resolution grids. 

\begin{figure}[ht]
     \begin{center}
      \setlength{\unitlength}{1cm}
       \begin{picture}( 5.,6.)           
        \put(-4.5,-1.5){\includegraphics{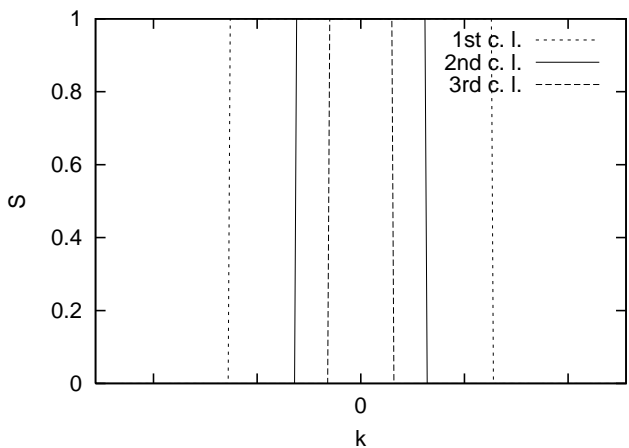}}   
       \end{picture}
 \caption[]{\label{F_ideal} The ideal function $S^l(k)$ defined in Eq.~\ref{sdef}
on three resolution levels denoted by first coarse level, second coarse 
level and third coarse level. On the zeroth original level the function is identical 
to one over the entire interval.
It gives a perfect dyadic decomposition of the Fourier spectrum.}
      \end{center}
     \end{figure}

Let us now postulate ideal restriction and prolongation operators and 
discuss their properties. As follows from the discussion above, they 
should provide for a dyadic decomposition of the Fourier space. 
Consequently the restriction operator would have to be a perfect 
low pass filter for the lower half of the spectrum (in the 1-dim case, 
in the 3-dim case only 1/8 would survive). We will refer to this 
property in following as frequency separation property.
The degree of perfectness can be quantified by the $k$ dependent function
\begin{equation}
 S^l(k) = \sqrt{\sum_i ( s^l_i )^2 / N} \label{sdef}
\end{equation}
where $N$ is the number of grid points on resolution level $l$.
The $k$ dependence enters through the requirement that 
the signal $s^{Lmax}_i$ on the finest resolution level is a pure harmonic, 
\begin{equation}
s^{Lmax}_i = \exp(I 2 \pi k i / N)  \label{shar}
\end{equation}
The function $S^l(k)$ for a perfect 
restriction operator is shown in Fig~\ref{F_ideal}.
Such ideal grid transfer operators have to 
satisfy a second property, that will be baptized the identity property. 
The prolongation operator has to bring back 
exactly onto the finer grid the long wavelength associated with the 
coarser grid. This can only be true if prolongation followed by 
a restriction gives the identity. A third desirable property would be 
that the coarse charge density represents as faithfully as possible the 
significant features of the original charge density. In particular the 
coarse charge density should have the same multipoles and most importantly 
the same monopole. The conservation of the monopole just means that 
the total charge is identical on all grid levels. This third property 
will be called the multipole conservation property in the following. 

With the ideal grid transfer operators, the solution of Poisson's 
equation would be a divide and conquer approach in Fourier space and 
it could be done with a single V cycle with a moderate number 
of GS relaxations on each resolution level. In contrast to a solution 
in a plane wave basis the division would not be into single Fourier 
components but into dyadic parts of the Fourier spectrum. For the case 
of our postulated ideal grid transfer operators it also would not matter 
whether the GS relaxations are applied when going up or going down, only 
the total number of GS relaxations on each level would count.

To establish the relation between multigrid grid transfer operators 
and wavelet theory, let us point out a formal similarity.
For vanishing $d$ coefficients, the 
wavelet analysis step is given by (Eq. 26 of ref.~\cite{mywvltbook}) 
\begin{equation} \label{genrestriction}
 s_{i}^{2h} = \sum_{j=-m}^{m} \tilde{h}_{j} s_{j + 2 i}^{h}  \label{forward}
\end{equation}
and is formally identical to a restriction operation. 
A wavelet synthesis step for the $s$ coefficients is given by
(formula 27 of ref.~\cite{mywvltbook}) 
\begin{eqnarray}
s_{2 i   }^{h} &=& \sum_{j=-m/2}^{m/2} h_{2 j  } \: s_{i-j}^{2h} 
 \label{genprolongation} \\
s_{2 i+1 }^{h} &=& \sum_{j=-m/2}^{m/2} h_{2 j+1} \: s_{i-j}^{2h} \nonumber \: .
\end{eqnarray}
and is formally identical to a prolongation operation. 
One can now for example easily see that 
the injection scheme for the restriction 
and linear interpolation for the prolongation part 
corresponds to a wavelet analysis and synthesis steps 
for 1st order interpolating wavelets. 
Using the values of the filters $\tilde{h}$ and $h$ for interpolating  
wavelets we obtain
\begin{equation} \label{injection} 
 s_{i}^{2h} =  s_{2 i}^{h}  
\end{equation}
and
\begin{eqnarray}
s_{2 i   }^{h} &=& s_{i}^{2h} \label{interpol} \\
s_{2 i+1 }^{h} &=&  \frac{1}{2} \: s_{i}^{2h} + \frac{1}{2} \: s_{i+1}^{2h} \nonumber \: .
\end{eqnarray}
which is the standard injection and interpolation. As a consequence of 
the fact that it can be considered as a wavelet forward and backward 
transformation, the combination 
of injection and interpolation satisfy the identity property of 
our ideal grid transfer operator pair, namely that 
applying a restriction onto a prolongation gives the identity.

Usually injection is replaced by the full weighting scheme, 
\begin{equation} \label{fullweight}
 s_{i}^{2h} =  \frac{1}{4} \: s_{2 i - 1}^{h} + 
  \frac{1}{2} \:  s_{2 i}^{h} + \frac{1}{4} \: s_{2 i + 1}^{h}   \: .
\end{equation}
This scheme has the advantage that it conserves averages, i.e it 
satisfies the monopole part of the multipole conservation property of an ideal 
restriction operator. Applying it 
to a charge density thus ensures that the total charge is 
the same on any grid level. Trying to put the full weighting 
scheme into the wavelet theory framework gives a filter 
$\tilde{h}$ with nonzero values of 
$\tilde{h}_{-1}=\frac{1}{4}$, $\tilde{h}_{0}=\frac{1}{2}$, 
$\tilde{h}_{1}=\frac{1}{4}$ This filter $\tilde{h}$ does not satisfy 
the orthogonality relations of wavelet theory 
(formula 8 of ref.~\cite{mywvltbook}) with the $h$ filter 
corresponding to linear interpolation. Hence a
prolongation followed by a restriction does not give the identity.

A pair of similar restriction and prolongation operators that conserve 
averages can however be derived from wavelet theory. Instead of 
using interpolating wavelets we have to use lifted interpolating 
wavelets~\cite{schneider,sweldens}. In this way we can obtain both 
properties, average conservation and the identity for a 
prolongation restriction sequence.  Using the filters derived in 
ref.~\cite{mywvltbook} we obtain
\begin{equation} \label{myrestriction}
s_{i}^{2h} = - \frac{1}{8} \: s_{2 i - 2}^{h}  
             + \frac{1}{4} \: s_{2 i - 1}^{h}  
             + \frac{3}{4} \: s_{2 i}^{h} 
             + \frac{1}{4} \: s_{2 i + 1}^{h}  
             - \frac{1}{8} \: s_{2 i + 2}^{h}  
\end{equation} 
\begin{eqnarray}
s_{2 i   }^{h} &=& s_{i}^{2h} \label{myprolongation} \\
s_{2 i+1 }^{h} &=&  \frac{1}{2} \: s_{i}^{2h} + \frac{1}{2} \: s_{i+1}^{2h} \nonumber \: .
\end{eqnarray}

\begin{figure}[p]
     \begin{center}
      \setlength{\unitlength}{1cm}
       \begin{picture}( 5.,18.)           
        \put(-7.5,11.){\includegraphics{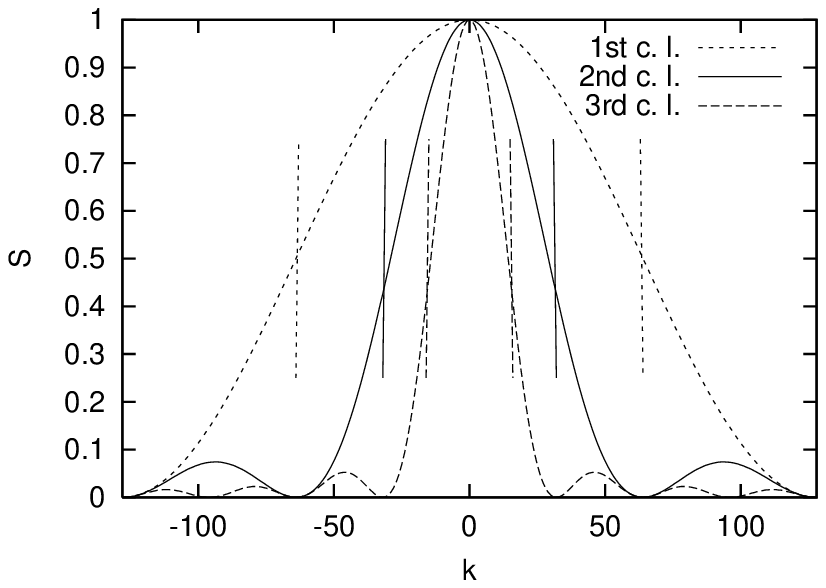}}   
        \put(1.5,11.){\includegraphics{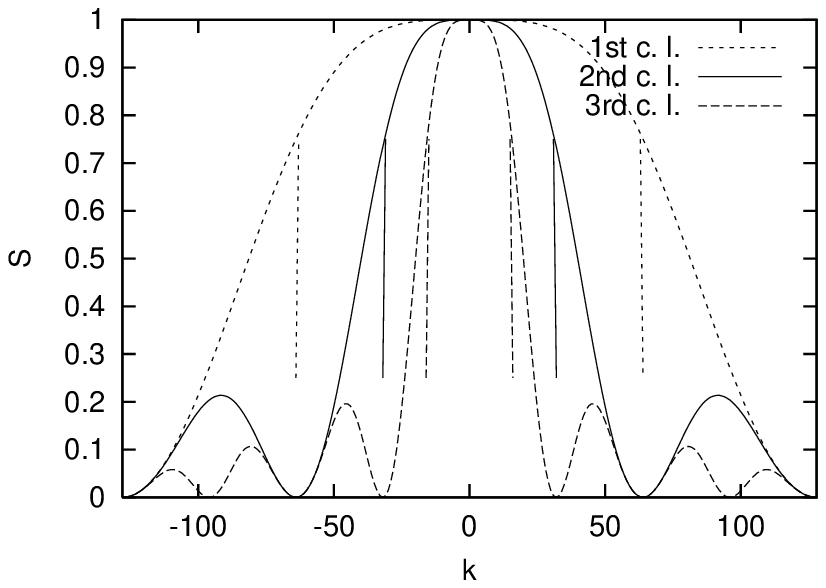}}   
        \put(-7.5,5.){\includegraphics{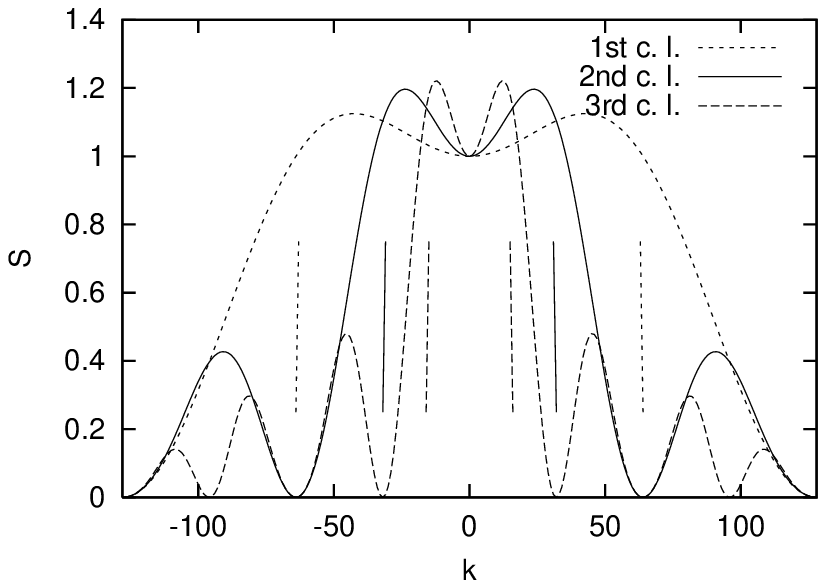}}   
        \put(1.5,5.){\includegraphics{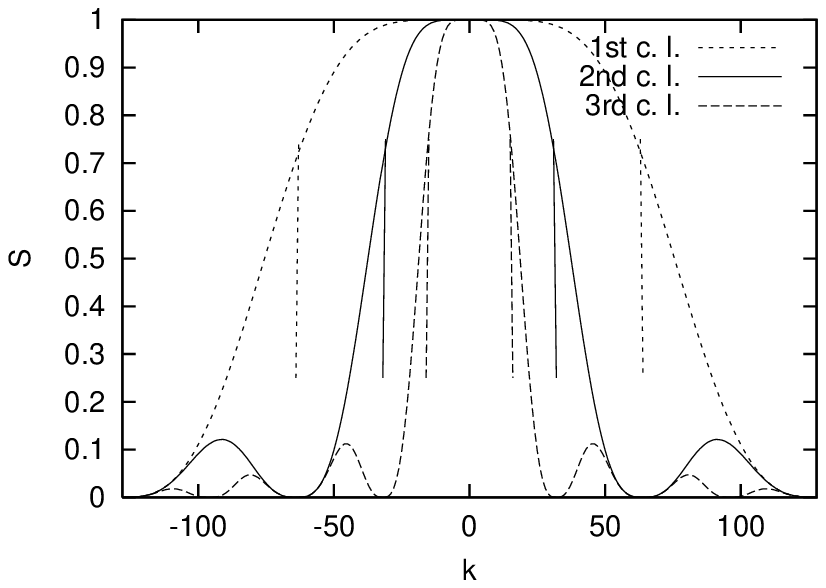}}   
        \put(-7.5,-1.){\includegraphics{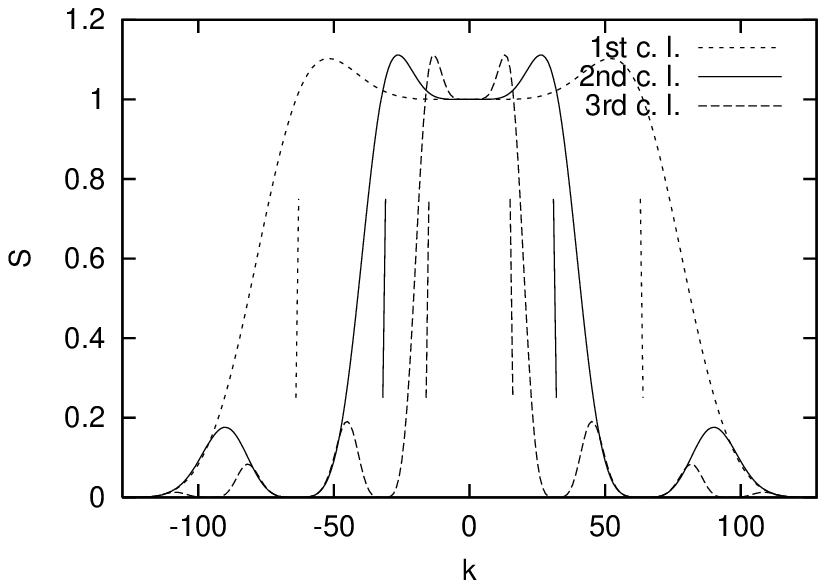}}   
        \put(1.5,-1.){\includegraphics{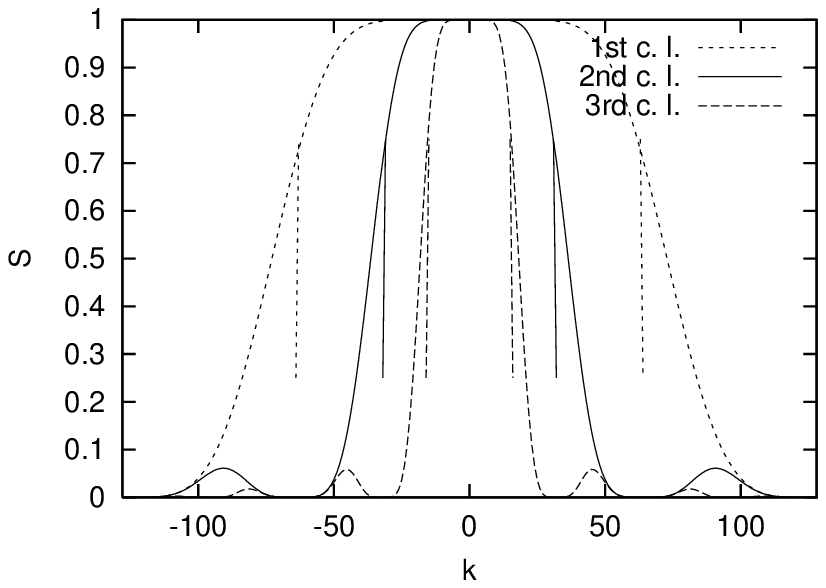}}   
       \end{picture}
 \caption[]{\label{fan} The function $S^l(k)$ defined in Eq.~\ref{sdef} 
 on the coarser resolution levels $Lmax-1$, $Lmax-2$ and $Lmax-3$ 
(corresponding to grid spacings of 2h, 4h and 8h if the finest resolution is h) 
for several restriction operators: Top left, Full weighting; middle left, 2nd order 
lifted wavelets, bottom left 6th order twofold lifted wavelets; top right 
half and half mixture between Full Weighting and 2nd order lifted wavelets; 
middle right, 6th order Daubechies; bottom right, 10th order Daubechies. $S$ was 
calculated numerically for an initial data set of 256 points. Hence the allowed 
values of $k$ in Eq.~\ref{shar} range form -128 to 127. The lower half, quarter and 
eight of the spectrum where the ideal function would switch between the values of 
0 and 1 are denoted by vertical lines. }
      \end{center}
     \end{figure}

Let us finally discuss the first property of our postulated ideal restriction 
operator, namely that it is a perfect low pass filter. Obviously any finite 
length filter can only be an approximate perfect low pass filter. Fig~\ref{fan} 
shows the $S$ function for several grid transfer operators. One clearly sees that 
the Full Weighting operator is a poor low pass filter, the filter derived from 
second degree lifted wavelets is already better and the filters 
obtained from 10th degree Daubechies wavelets and twofold lifted 
6th order interpolating wavelets are best. The Daubechies 6th degree filter 
is intermediate and of nearly identical quality as the one that is a mixture 
of the Full Weighting and second degree lifted wavelet filters. In contrast to the former, 
the later does however not fulfill the identity property.

The degree of perfectness for the frequency separation 
is also related to the multipole conservation property of our 
postulated ideal grid transfer operators. As one sees from Fig.~\ref{fan}
filters which correspond to wavelet families with many vanishing are 
closer to being ideal for frequency separation 
than those with few vanishing moments. At the same time the number of 
vanishing moments determines how many multipoles are conserved when the charge 
density is brought onto the coarser grids. 

\begin{figure}[p]
     \begin{center}
      \setlength{\unitlength}{1cm}
       \begin{picture}( 5.,18.)           
        \put(-7.5,11.){\includegraphics{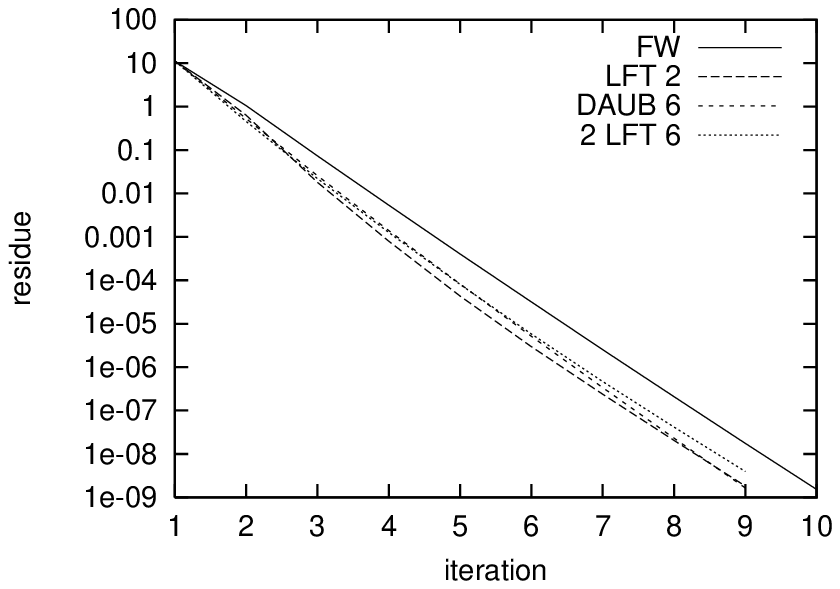}}   
        \put(1.5,11.){\includegraphics{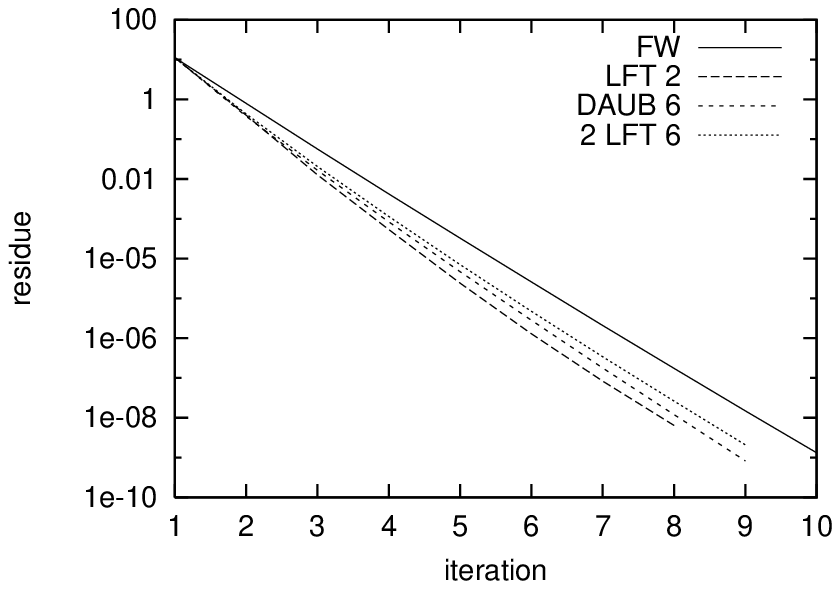}}   
        \put(-7.5,5.){\includegraphics{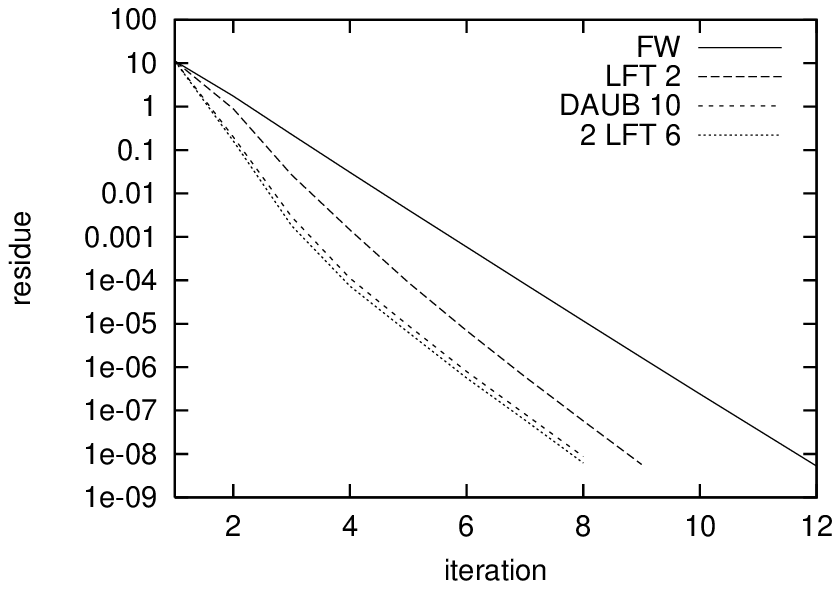}}   
        \put(1.5,5.){\includegraphics{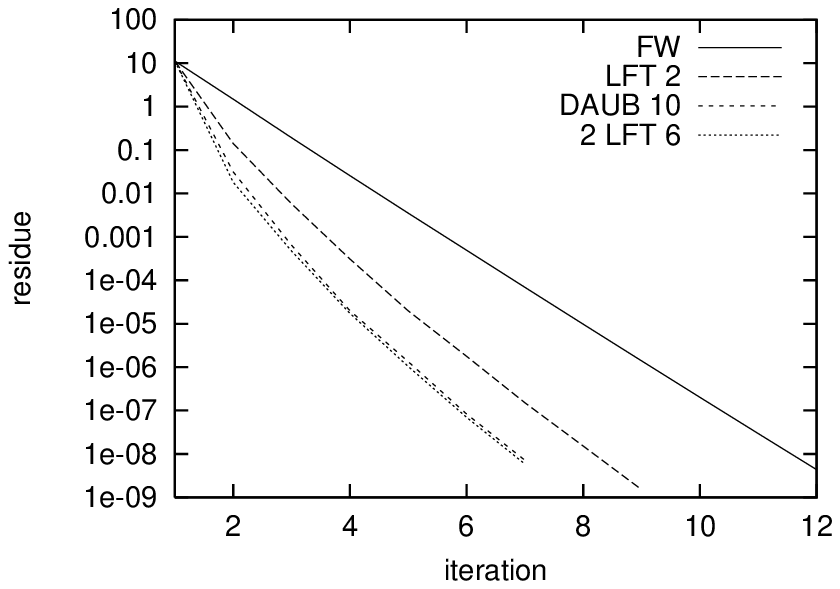}}   
        \put(-7.5,-1.){\includegraphics{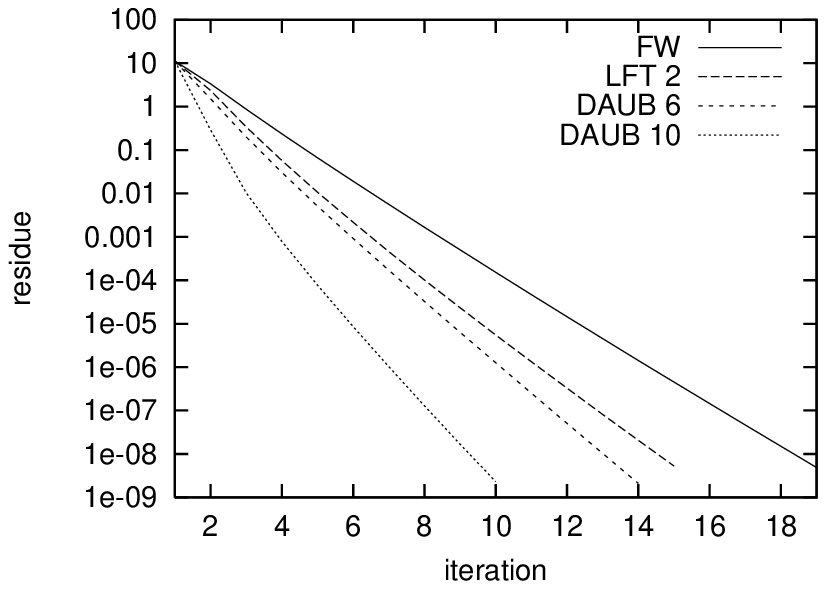}}   
        \put(1.5,-1.){\includegraphics{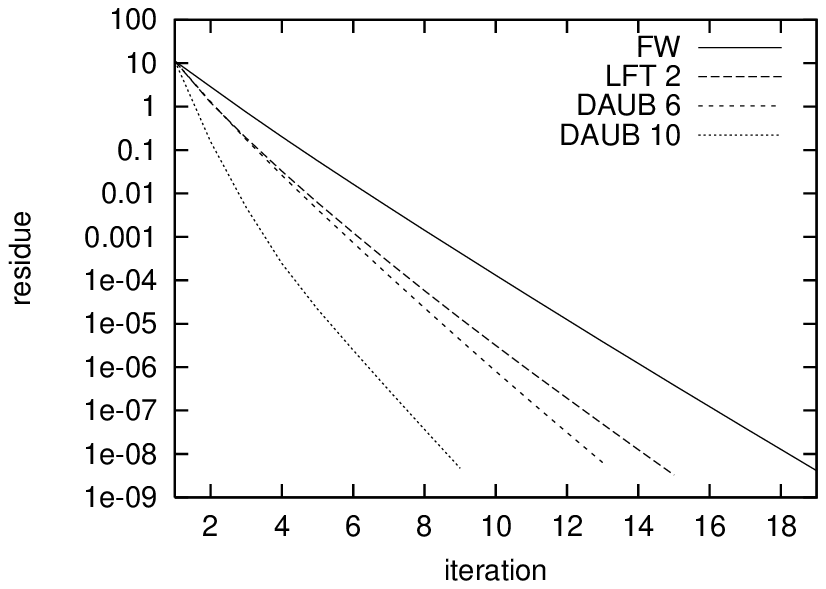}}   
       \end{picture}
 \caption[]{\label{comp} The convergence rate of a sequence of V cycles 
  (left hand side) and halfway V cycles (right hand side). In the 
  upper two plots Poisson's equation was discretized by second order 
  finite differences, In the middle two plots by 6-th order finite 
  differences and in the lower two plots by 6-th order and 10th order 
 (for the case of transfer operators based on DAUB 10) interpolating 
  scaling functions. Shown are results for the Full Weighting scheme 
 (FW) second order lifted wavelets (LFT 2), twofold lifted 6-th order 
 wavelets (2 LFT 6) and 6-th and 10-th order Daubechies wavelets. In the case of 
 ordinary V cycles 2 GS relaxations were done on the finest level 
 both when going up and coming back down, in the case of the halfway V cycle 4 
 GS relaxation were done on the finest level. }
      \end{center}
     \end{figure}

The right panel of Fig.~\ref{comp} shows the convergence rate of a 
sequence of V cycles 
for the full weighting/interpolation (Eq.~\ref{fullweight},\ref{interpol}) 
scheme and various wavelet based schemes, namely the scheme 
obtained from second order lifted wavelets 
(Eq.~\ref{myrestriction},\ref{myprolongation}), the corresponding scheme, 
but obtained from twofold lifted 6-th order wavelets as well 
as schemes obtained from 6th and 10th order Daubechies wavelets. 
The numerical values for the filters are listed in the Appendix. 
One can observe a clear correlation between the convergence rate and the 
the degree of perfectness of the $S$ function. A high degree of perfectness 
is particularly useful in connection with high order discretizations of 
the Laplacian. Most of the filters of the grid transfer operators are longer 
than the standard Full Weighting filter, which just has 3 elements. The 
lifted 2nd order interpolating wavelet restriction filter has for instance 
5 elements and the 6-th degree Daubechies filter 6 elements. This does however 
not lead to a substantial increase of the CPU time. This comes from the fact 
that on modern computers the transfer of the data into the cache is the 
most time consuming part. How many numerical operations are then performed 
on these data residing in cache has only a minor influence on the timing. 
The new wavelet based schemes for restriction and prolongation 
are therefore more efficient than the Full Weighting 
scheme, both for finite difference discretizations and 
scaling function basis sets. It is also obvious that the multigrid approach for 
scaling/wavelet function basis sets is more efficient than the diagonal 
preconditioning approach. 

The identity property for a restriction prolongation operator pair was only 
necessary for the case of operators where the restriction part is a perfect 
low pass filter. One might therefore wonder how useful it is in the context 
of the only nearly perfect filters. The numerical experience suggests that 
it is nevertheless a useful property. One can for example compare the 
convergence rates using either the 6-th order Daubechies filters or the 
filter that is the average of Full Weighting and lifted 2nd order wavelet 
filters. Fig.~\ref{fan} shows that their restriction parts have very 
similar $S$ functions. Nevertheless we always found a better convergence rate 
with the Daubechies filter which satisfies the identity property.

For the 2-dimensional Poisson equation it has been shown, that the convergence rate 
compared to the standard Full Weighting scheme can be improved by tailoring 
grid transfer operators for the relaxation scheme used~\cite{achi1}. The 
theoretical foundations for this is furnished by local Fourier analysis~\cite{achi23}. 
The same approach could certainly also be used for the 3-dimensional case considered 
here. It is to be expected that the grid transfer operators found by such an 
optimization would be very close to the ones that we have obtained from 
wavelet theory. 

\begin{figure}[ht]
     \begin{center}
      \setlength{\unitlength}{1cm}
       \begin{picture}( 5.,5.5)           
        \put(-8.,-11.){\includegraphics{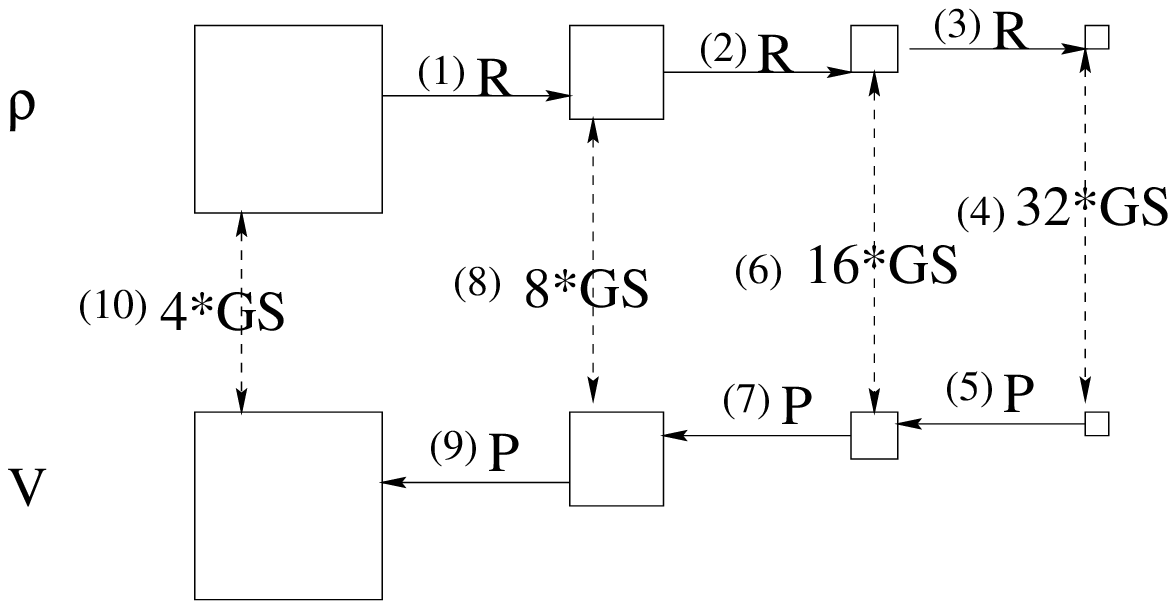}}   
       \end{picture}
 \caption[]{\label{mg} Schematic representation of a halfway V
cycle as described in the text. The abbreviations are the same as in 
Fig.~\ref{mgv}. }
      \end{center}
     \end{figure}

The main justification for the relaxations in the upper part of the 
traditional multigrid algorithm shown in Fig.~\ref{mgv} is to eliminate 
the high frequencies. This can however be done directly by fast wavelet 
transformations based on wavelets that have good localization properties in 
frequency space such as lifted interpolating wavelets. As a consequence 
the traditional multigrid algorithms can be simplified considerably 
as shown in Fig.~\ref{mg}. Using wavelet based restriction and 
prolongation operators we can completely eliminate 
the GS relaxation in the first part of the V cycle where we go 
from the fine grid to the coarsest grid. We baptize such a simplified 
V cycle a halfway V cycle. The numerical results, obtained with the halfway 
V cycle, shown in the right hand plots of Fig.~\ref{comp}, demonstrate 
that the convergence is slightly faster than for the traditional 
multigrid algorithm based on the same restriction and prolongation scheme. 
In addition one step is faster. It is not necessary to calculate the 
residue after the GS relaxations. Otherwise the number of GS relaxations 
and restrictions/prolongations is identical in the full and halfway 
V cycle. On purpose no CPU times are given in this 
context because optimization of certain routines~\cite{mybook} can entirely 
change these timings. 
Because the residue is never calculated in the halfway V cycle, 
the memory requirements are also reduced.

\begin{figure}[ht]
     \begin{center}
      \setlength{\unitlength}{1cm}
       \begin{picture}( 5.,6.)           
        \put(-7.5,-1.5){\includegraphics{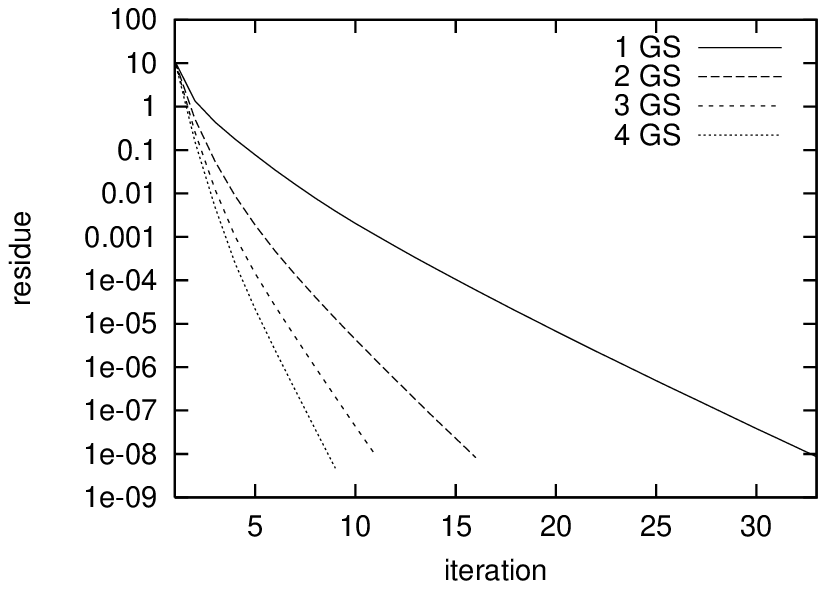}}   
        \put(1.5,-1.5){\includegraphics{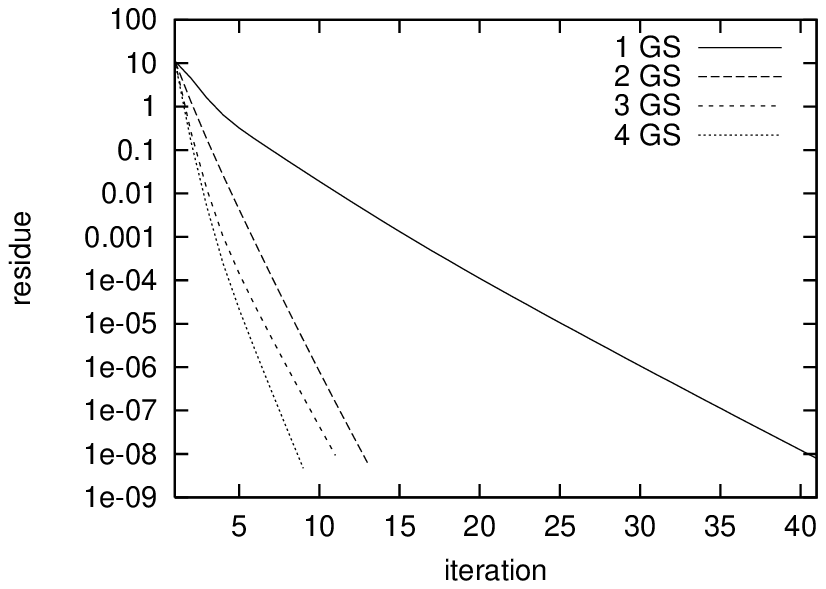}}   
       \end{picture}
 \caption[]{\label{num_gs} The convergence rate for halfway V cycles 
 with 4, 3, 2 and 1 GS relaxation on the finest grid level. In the left panel 
 2nd order finite differences were used, in the right panel 6th order finite 
 differences. }
      \end{center}
     \end{figure}

The number of GS relaxations in the halfway V cycle was chosen to be 
4 in order to allow for an unbiased comparison with the 
traditional V cycle where also 4 GS relaxations were done on the 
finest grid level. For optimal overall efficiency putting the 
number of GS relaxation to 3 is usually best, with the values of 2 and 
4 leading to a modest increase in the computing time. The convergence 
rate of halfway V cycles as a function of the number of GS relaxations 
on the finest grid level is shown in Fig~\ref{num_gs}. 

In all the previous examples we specified the number of GS relaxations on the 
finest grid level. On the coarser grid levels the number of iterations 
was allowed to increase by a factor of two per grid level. In this way 
it was practically always possible to find the exact solution on the most 
coarse grid. In addition we found that this trick slightly reduces the number 
of iterations and the total CPU time. The overall behavior of all the 
different methods is however identical when the number of GS relaxation is 
constant on each grid level.

\section{Conclusions}
Our results demonstrate that halfway V cycles with the restriction and 
prolongation steps based on wavelet theory are the most efficient approach 
for the solution of the 3-dimensional Poisson's equation. It is most efficient both 
for finite difference discretizations and for the case where 
scaling functions or wavelets are used as basis functions. We expect 
that the approach should also be the most efficient one in 
connection with finite elements. It is essential that the wavelet 
family used for the derivation of the restriction and prolongation 
schemes has at least one vanishing moment and conserves thus average 
quantities on the various grid levels. Wavelet families with more 
vanishing moments do not lead an appreciable increase of the convergence 
rate compared to the case of one vanishing moment for low order 
discretizations of Poisson's equation, but lead a modest further 
increase for high order discretizations. In the case where a wavelet family 
was used to discretize the Laplace operator, it is best to use the same 
wavelet family to construct the grid transfer operators.
In addition to increased efficiency of the proposed halfway V cycle 
in terms of the CPU time, it is also simpler 
than the standard V cycle. This makes not only programming easier, 
but also reduces the memory requirements.

\section{Acknowledgments}
I was fortunate to have several discussion with Achi Brandt about this work.
His great insight on multigrid methods, that he was willing to share with me,  
helped a lot to improve the manuscript. I thank him very much for 
his interest and advice. 

\section{Appendix}

Filter for twofold lifted 6th order interpolating wavelets~\cite{lift}: 
\newline
$h_1$=75/128,$h_3$=-25/256,$h_5$=3/256, 
\newline
$\tilde{h}_0$=2721/4096, 
$\tilde{h}_1$=9/32, $\tilde{h}_2$=-243/2048, $\tilde{h}_3$=-1/32,
\newline
$\tilde{h}_4$=87/2048, $\tilde{h}_6$=-13/2048, $\tilde{h}_8$=3/8192 . 
\newline
The values for negative indices follow from the symmetry $h_i = h_{-i}$ \
and $\tilde{h}_{i}= \tilde{h}_{-i}$.

Filters for 6-th order Daubechies wavelets~\cite{daub}: 
\newline
$h_{-2}$=0.3326705529500826159985d0, $h_{-1}$=0.8068915093110925764944d0, 
\newline
$h_0$=0.4598775021184915700951d0, $h_1$=-0.1350110200102545886963d0, 
\newline
$h_2$=-0.0854412738820266616928d0, $h_3$= 0.0352262918857095366027d0 .
\newline
$\tilde{h}_{i}= {h}_{i}$.

Filters for 10-th order extremal Daubechies wavelets~\cite{daub}: 
\newline
$h_{-4}$=.1601023979741929d0, $h_{-3}$=.6038292697971897d0, 
\newline
$h_{-2}$=.7243085284377729d0, $h_{-1}$=.1384281459013207d0, 
\newline
$h_0$=-.2422948870663820d0, $h_1$=-.0322448695846384d0,
\newline
$h_2$=.0775714938400457d0, $h_3$= -.0062414902127983d0,
\newline
$h_4$=-.0125807519990820d0, $h_5$=.0033357252854738d0 .
\newline
$\tilde{h}_{i}= {h}_{i}$.

\end{document}